\documentclass[11pt,letterpaper]{article}
\usepackage{color}
\usepackage{amsmath}    
\usepackage{amssymb}
\usepackage{graphicx} 
\DeclareGraphicsExtensions{.png,.jpg,.mps,.gif,.bmp}
\usepackage{float}
\usepackage{authblk}
\usepackage[letterpaper]{geometry}
\usepackage[final]{pdfpages}

\begin{document}     

\title{A Mechano-Chemical model for tumors growth}

\author[1]{Cristian C. P{\'e}rez {\'A}guila}
\author[2]{Maura C{\'a}rdenas G.}
\author[1]{J. Fernando Rojas} 
\affil[1]{Facultad de Ciencias F\'isico Matem\'aticas.\\ Benem\'erita Universidad Aut\'onoma de Puebla,\\Av. San Claudio y 18 sur, Ciudad Universitaria, Col. San Manuel.\\ C. P. 72570. Puebla, M\'exico.}
\affil[2]{Facultad de Medicina. Benem\'erita Universidad Aut\'onoma de Puebla\\ Calle 13 sur 2702, Col. Los Volcanes, C. P. 72420 Puebla, M\'exico}
  

\date{}  
\maketitle  
  
\begin{abstract}
In this paper we present a study of local dynamics of the growth of
  cancer tumor and healthy cells considering the presence of nutrients in the system. We also analyze the evolution of system if we take indirectly into account the level of
  alkalinity (pH) in the system which shows that influences in tumor growth. The model consists in a set of differential equations of first order that involves a mechanical model added by a pair of differential equations for local oxygen and glucose.
\end{abstract}




\section{Introduction}
\label{S:1}
Cancer is a disease of uncontrolled cell proliferation \cite{michalopoulou}. These cells exhibit metabolic alterations \cite{anastasiou} that distinguish them from healthy tissues and make their metabolic processes susceptible to pharmacological targeting \cite{anastasiou}. Successful cell proliferation is dependent on a profound remodeling of cellular metabolism, required to support the biosynthetic needs of a growing cell \cite{pavlova}. In the literature we find that the conditions of the tumor microenvironment are important in the metabolism of the tumoral cells \cite{anastasiou}, this allows cells to use available nutrients such as glucose and glutamine \cite{anastasiou, michalopoulou, pavlova, swietach} for its survival and propagation. Some tumours show increased glucose uptake, that can be promoted by gene expression or oncogenic  signalling changes \cite{anastasiou, pavlova}.
Signaling through oncogenes allows cancer cells to perpetually receive pro-growth stimuli, which for normal tissues are only transient in nature. As a consequence, cancer metabolism is constitutively geared toward supporting proliferation \cite{pavlova}.

Some types of cancer show the nutrients are insufficient. In  recent work has shown that cancer cells are able to overcome this nutrient insufficiency by scavenging alternative substrates as proteins and lipids \cite{michalopoulou}.\\

Another problem that arises is the reduction of oxygen supply or hypoxia. Lack of sufficient oxygen has a profound effect on cellular metabolism as it inhibits those biochemical reactions in which oxygen is consumed \cite{pavlova}.
One of the most important phenotypic changes is the cancer cell’s ability to change the extracellular acidity due to an altered glycolysis pathway. The ability to expel protons from cells and, therefore, change the extracellular pH provides an advantage to cancer cells \cite{walsh}.\\

In recent works it is clear that pH plays an important role in the survival mechanisms of cancer cells \cite{walsh, pavlova, anastasiou, swietach}. The acidic tumour microenvironment has also been associated with the degree of cancer aggressiveness. It has been reported that the lower the pH surrounding the cell, the more likely is the chance of malignancies with higher degree of invasiveness \cite{walsh}. A recent therapeutic approach takes advantage of the altered acidity of the tumour microenvironment by using proton pump inhibitors (PPIs) to block the hydrogen transport out of the cell. The alteration of the extracellular pH kills tumour cells, reverses drug resistance, and reduces cancer metastasis. Targeting tumour pH can be therefore considered a therapeutic strategy \cite{walsh}.

\section{Components of system.}

A tumor is an example of a complex system composed of cells. Each of the
the reproduced cells follows a set of rules and responds to local interactions with
other cells either from the tumor or with healthy cells.\\
\\
\textbf{Extracellular matrix.}\\
The Extracellular matrix (ECM) represents a network that includes to all
organs, tissues and cells of the body, in this medium the cells survive,
multiply and perform their vital functions {\cite{matriz}}.\\
The balance of extracellular matrix is maintained by a regulation between
synthesis, training and reshuffle. Also the extracellular matrix provides oxygen and nutrients to the cell and the elimination of $CO_2$ and toxins under
normal conditions {\cite{matriz}}.\\
The extracellular matrix of animals need to reorganize and also need a constant regeneration through of the degradation of components and the production of
new components by cells. The degradation of ECM causes enzymes production such as
metalloproteinases {\cite{matex}}.\\
\\
\textbf{Extracellular liquid.}\\
The extracellular liquid is body fluid not contained in cells. Often it is secreted by cells for an suitable environment. It is transported across the body in two stages, the first is the movement of blood by \textit{circulatory
system}; the second is movement between blood capillaries and cells.\\
When blood flow through the capillaries, a large portion of the liquid containing is diffused into and out of interstitial liquid that found between cells, allowing the continuous exchange of substances between the cells and interstitial liquid and between the interstitial liquid and the blood.\\
\\
\textbf{Source of nutrients in the extracellular liquid.}\\
$\bullet$ The \textit{respiratory system}, it provides oxygen to the body
and removes carbon dioxide.\\
$\bullet$ The \textit{digestive system}, digests food and absorbs nutrients including carbohydrates, fatty acids, and amino acids, for to bring into the
extracellular liquid.\\
\\
\textbf{The blood.}\\
Transports nutrients, electrolytes, hormones, vitamins, antibodies, heat and
oxygen, its composed of plasma, red and white blood cells, platelets. The
principal function of red blood cells, transports oxygen and carbon dioxide.
Glucose is a sugar which comes of food we eat, this glucose is stored in the
body, is the main source of energy for body cells and reaches every cells
through the bloodstream.\\
\\

\section{Mathematical model}

We consider a mixture in which the essential components are cells, extracellular matrix, and extracellular liquid with dissolved chemicals. Hypotheses that are assumed are as follows:\\
\\
$\bullet$ The components of the extracellular matrix form a complex network and move together.\\
$\bullet$ Cells responds mechanically to compression from cells around, regardless of the type of cell.\\
$\bullet$ The presence of nutrients in the mixture is essential for growth and
vital functions of cells \cite{metab}.\\
\\
For this model is defined $\phi \in [0, 1]$ as the volume occupied by cells and either $m \in [0, 1]$ as the volume occupied by extracellular matrix. The mixture is saturated if the remaining space is filled with extracellular liquid then we define $l \in [0, 1]$ as the volume occupied by extracellular liquid. Considering normalized amounts for the sample volume we have:
\begin{equation}
  \phi + m + l = 1
\end{equation}
In the cellular component of tissue, the differential equation that describes the change in mass of the cellular component is
\begin{equation}
  \frac{\partial \phi}{\partial t} + \nabla \cdot (\phi
  \overrightarrow{v}_\phi) = \Gamma_{\phi}
\end{equation}
which considers that cell density is constant ({\cite{astanin}},
{\cite{tosin}}, {\cite{param}}). Here $\Gamma_{\phi}$ corresponds to
rapid growth or death for cell mass, $\overrightarrow{v}_\phi$ is the cell
rate and the term $\nabla \cdot (\phi \overrightarrow{v}_\phi)$ it is
associated with the flow of cell component\\
In the same way, the corresponding equations for the components of the mixture
\begin{equation}
  \frac{\partial \phi}{\partial t} + \nabla \cdot (\phi
  \overrightarrow{v}_\phi) = \Gamma_{\phi}
\end{equation}
\begin{equation}
  \frac{\partial m}{\partial t} + \nabla \cdot (m \overrightarrow{v}_m) =
  \Gamma_m
\end{equation}
\begin{equation}
  \frac{\partial l}{\partial t} + \nabla \cdot (l \overrightarrow{v}_l) =
  \Gamma_l.
\end{equation}
For the system of interest it is needed to distinguish different types of cells such as tumor cells, endothelial, epithelial, fibroblasts, etc. However, for simplicity, in the cell mass only we only distinguish two subpopulations that correspond to \textit{healthy} cells and tumor cells. In the extracellular matrix components all are considered as a single assembly.
\begin{equation}
  \frac{\partial \phi_i}{\partial t} + \nabla \cdot (\phi_i
  \overrightarrow{v}_{\phi_i}) = \Gamma_{\phi_i}, \quad i = n, t,
\end{equation}
and considering the result obtained in {\cite{astanin}} and {\cite{tosin}}, cell
velocity is:
\begin{equation}
  \overrightarrow{v_i} = - K_{im}  \left( 1 - \frac{\sigma_{im}}{| \nabla
  \cdot (\phi_i  \mathbf{T}_{\phi}) |} \right) \nabla (\phi_i 
\mathbf{T}_{\phi})  \qquad i = n, t.
\end{equation}
In this equation it is considered $\sigma_{im} = 0$ ({\cite{param}}), that is associated with frictional forces and it\'{}s expected to depends on adhesion mechanisms of volume occupied by cells. Also it is supposed that the coefficient of mobility $K_{im} = K_i$ is constant, while the term $\mathbf{T}_{\phi}$ expresses stress to which the cells are subjected.\\
Now if we divide the sample volume in small cubes in which only there is a cell population healthy or tumor, then $\phi_i$ is constant in each cube, and in this way the cell velocity can be expressed as
\begin{equation}
  \overrightarrow{v}_i = - K_i \nabla (\phi_i  \mathbf{T}_{\phi})  \qquad i
  = n, t
\end{equation}
where, the term associated with the flow can be expressed as follows
\begin{equation}
  \nabla \cdot (\phi_i \overrightarrow{v}_{\phi_i}) = - \phi_i K_i \nabla^2 
  (\phi_i  \mathbf{T}_{\phi}).
\end{equation}
It is necessary an equation that describes the response of cells and extracellular
matrix to stress, this equation is constructed assuming that they behaves like an
elastic fluid ({\cite{astanin}}, {\cite{nutri}}, {\cite{kundu}},
{\cite{tosin}}), so we consider
\begin{equation}
  \mathbf{T}_{\phi} = - \Sigma I,
\end{equation}
and for the function $\Sigma$, we use the expression proposed by Chaplain et. al. {\cite{param}}
\begin{equation}
  \Sigma (\psi) = \left( \frac{2 - \psi_0}{1 - \psi_0} \right) \left(
  \frac{\psi - \psi_0}{1 - \psi} \right) ,
\end{equation}
where $\psi = \phi_n + \phi_t + m$, measure indirectly, the free space
available locally and it can be use for calculate the stress exerted by the
environment on cellular matter. Also $\psi_0 \in (0, 1)$, is identified as
the volume free of stress.

\subsection{Term of growth or death cell.}

Cells live in a crowd environment and they perceive the presence of other cells. This fact is fundamental in controlling cell concentration and implies to stop mitosis (cellular division) when the volume exceeds a threshold, so the behavior of cells in terms of growth and movement depends crucially on how they perceive the presence of other cells and its signal is translated.\\

For the terms of growth, we consider ({\cite{astanin}}, {\cite{param}},
{\cite{tosin}}), the next
\begin{equation}
  \Gamma_i = [\gamma_i H_{\sigma} (\psi - \psi_i) - \delta_i (\psi)] \phi_i,
  \quad i = n, t
\end{equation}
where $H_{\sigma}  (\psi - \psi_i)$ is a amendment to step function, and it is equal to 1 for $\psi$ smaller that the threshold value $\psi_i$ and is equal to zero for $\psi > \psi_i + \sigma$. Also it must be satisfied that $\psi_n < \psi_t$, and we consider constants $\delta_i$, and $\gamma_i > 0$ which represent coefficients of death and cell reproduction respectively. The parameter $\sigma > 0$ controls the transition between $H_{\sigma} (s) = 1$, and $H_{\sigma} (s) = 0$ ({\cite{param}}), therefore it controls the transition on/off in the cells reproduction, so the expression used in this work is
\begin{equation}
 H_{\sigma} (s) = \left\{ \begin{array}{l}
    1, \mbox{if } s \leq 0\\
    0, \mbox{if } s > \sigma\\
    1 - \frac{s}{\sigma}, \mbox{another case. }
  \end{array} \right.
\end{equation}

\subsection{Remodeling of the extracellular matrix.}

As mentioned on the biological part of extracellular matrix, now we consider a concentration of enzymes that degrades the ECM. The enzymes are produced by tumor or healthy cells{\cite{matex}} and so the process of remodelling can be described by the equation for $m$ as follows:
\begin{equation}
  \frac{\partial m}{\partial t} = \mu_n \phi_n \Sigma (\psi) + \mu_t \phi_t
  \Sigma (\psi) - \nu em.
\end{equation}
Here $\mu_n$ and $\mu_t$ corresponds to speed of production in the
extracellular matrix by healthy or tumor cells respectively, $\nu$ is the
degradation coefficient due to action of enzymes that degrade the matrix and
$e$ is the concentration of said enzymes {\cite{astanin}}.\\
\\
Now, we introduce an reaction-diffusion equation which aims to describe the
evolution in concentration of enzymes that degrade the extracellular matrix
\begin{equation}
  \frac{\partial e}{\partial t} = \kappa \nabla^2 e + \pi_n \phi_n \Sigma
  (\psi) + \pi_t \phi_t \Sigma (\psi) - \frac{e}{\tau}.
\end{equation}
In this equation $\pi_n$ and $\pi_t$ correspond to the rapid production of
enzymes that degrade the extracellular matrix produced by healthy and tumor
cells respectively; $\tau$ is the average lifetime and $\kappa$ the diffusion
coefficient of enzymes {\cite{astanin}}.

\subsection{Nutrients.}

In tumor growth, as well as vital functions of an healthy cell it is necessary the presence of dissolved chemicals in the liquid component of the mixture considered, such as nutrients, growth factors, etc. These chemicals are ``absorbed'' by cell populations to perform some vital functions such as proliferation, growth or intercellular communication {\cite{astanin}}. For the consideration of nutrients will assume that there is an source that supplies the nutrients, that is to say, there is a capillary in which blood flow is constant that transports nutrients.\\
\\
Considering only oxygen and glucose \cite{metab}, we propose the following
equations \cite{tosin}:
\begin{equation}
  \frac{\partial c}{\partial t} = D_c \nabla^2 c - \beta_n \phi_n c + f_c
\end{equation}
\begin{equation}
  \frac{\partial g}{\partial t} = D_g \nabla^2 g - \beta_t \phi_t g + f_g
\end{equation}
in which $c \in [0, 1]$ is the oxygen concentration and $g \in [0, 1]$ is the
glucose concentration with  $D_i$ ($i = c, g$) are the diffusion coefficients of nutrients and $\beta_j$ ($j = n, t$) corresponds to the absorption rate. Also $f_i$ represents the continuous supply of nutrients (i.e. a source). In the supply of nutrients must be met $f_c + f_g < 1$.\\
\\
Considering the nutrients that influence in the evolution of cell population, we assume that the rate of reproduction in cells is proportional to nutrients concentration, that would result replace $\gamma_n$ by $c \gamma_n$ ($\gamma_n \longrightarrow c \gamma_n$) for healthy cells and $\gamma_t \longrightarrow g \gamma_t$ for tumor cells, in the case of rate cell death is obtained $\delta_n \longrightarrow (1 - c) \delta_n$ and $\delta_t \longrightarrow (1 - g) \delta_t$ for healthy and tumor cells respectively, the reason for these substitutions is that for example if considered to healthy cells, with an optimal concentration of oxygen these are reproduced and the term of rate of death must decrease, this situation is described by the term $(1 - c)$. We must mention that competition is not considered between cell populations for nutrients. Thus, the complete mathematical model for the simulation of the system it is as follows:
\begin{equation}
  \left\{ \begin{array}{l}
    \frac{\partial \phi_n}{\partial t} = \phi_n K_n \nabla^2 (\phi_n \Sigma
    (\psi)) + c \gamma_n \phi_n H_{\sigma} (\psi - \psi_n) - (1 - c) \delta_n
    \phi_n\\
    \\
    \frac{\partial \phi_t}{\partial t} = \phi_t K_t \nabla^2 (\phi_t \Sigma
    (\psi)) + g \gamma_t \phi_t H_{\sigma} (\psi - \psi_t) - (1 - g) \delta_t
    \phi_t\\
    \\
    \frac{\partial m}{\partial t} = \mu_n \phi_n \Sigma (\psi) + \mu_t \phi_t
    \Sigma (\psi) - \nu em\\
    \\
    \frac{\partial e}{\partial t} = \kappa \nabla^2 e + \pi_n \phi_n \Sigma
    (\psi) + \pi_t \phi_t \Sigma (\psi) - \frac{e}{\tau}\\
    \\
    \frac{\partial c}{\partial t} = D \nabla^2 c - \beta_n \phi_n c + f_c\\
    \\
    \frac{\partial g}{\partial t} = D \nabla^2 g - \beta_t \phi_t g + f_g .
  \end{array} \right.
\end{equation}
which will be solved in an homogeneus space.

\section{Results}

In this section we show, the results of numerical solution for the system of
equations obtained, some parameters were obtained from {\cite{param}}, and we
propose some parameters for the simulations. We consider the cell population
is aproximately $50\%$, the extracellular matrix occupies a $20\%$ and
extracellular liquid a $30\%$ of the total volume.\\
We will take as reference the following parameters:

\begin{table}[h]
\centering
  \begin{tabular}{|c|c|c|c|c|c|}
    \hline
    $\phi_n = 0.45$ & $\phi_t = 0.005$ & $m = 0.2$ & $e = 0.3$ & $c = 0.25$ &
    $g = 0.16$\\
    \hline
  \end{tabular}
  \caption{Initial condition.}
\end{table}

\begin{table}[h]
\centering
  \begin{tabular}{|c|c|c|c|c|}
    \hline
    $K_n = 0.1$ & $\gamma_n = 0.746$ & $\psi_n = 0.6$ & $\delta_n = 0.1$ &
    $\mu_n = 0.1$\\
    \hline
    $\pi_n = 6000000$ & $\beta_n = 1.2$ & $K_t = 0.3$ & $\gamma_t = 0.97$ &
    $\psi_t = 0.8$\\
    \hline
    $\delta_t = 0.03$ & $\mu_t = 0.05$ & $\pi_t = 3000000$ & $\beta_t = 1.3$ &
    $\sigma = 0.2$\\
    \hline
    $\nu = 0.000016$ & $\kappa = 0.00000734$ & $\tau = 0.005$ & $D = 1.0$ &
    $\psi_0 = 0.75$\\
    \hline
    $f_c = 0.25$ & $f_g = 0.16$ &  &  & \\
    \hline
  \end{tabular}
  \caption{Set of parameters.}
\end{table}

Following graphs show the results of the local dynamics of the system
considered.\\
Figures 1 and 2 show dependence of cell populations in nutrient uptake, for
which is varied the absortion rate of tumor and healthy cells

\begin{figure}[h]
  \begin{center}
    \includegraphics[width=0.6\linewidth]{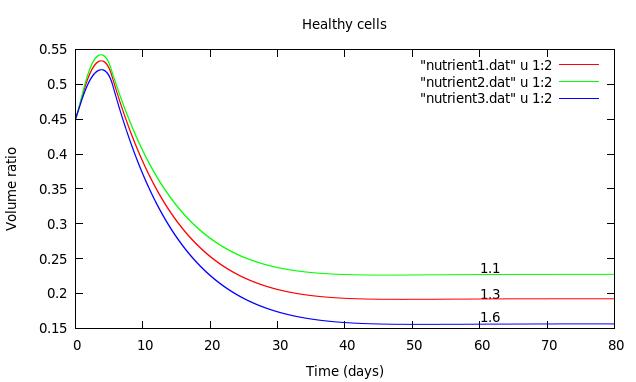} 
  \end{center}
  \caption{Evolution of healthy cells varying the rate of oxygen uptake.}
\end{figure}

\begin{figure}[h]
  \begin{center}
  \includegraphics[width=4.05in]{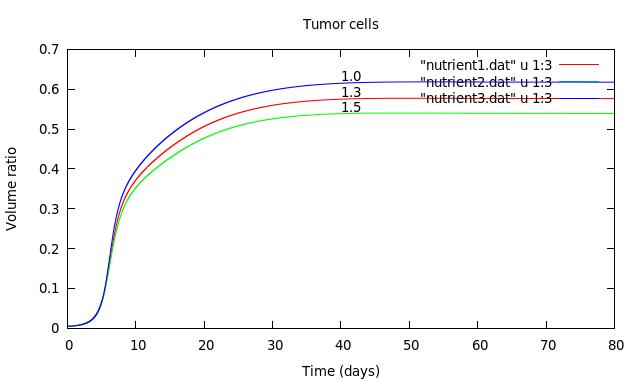}
  \end{center}
  \caption{Evolution of tumor cells varying the rate of glucose uptake.}
\end{figure}

In the graphs 3, 4, correspond to growth of cell population, varing the
concentration of oxygen, also we considered that $f_g = 0.1$.

\begin{figure}[h]
  \begin{center}
    \includegraphics[width=4.05in]{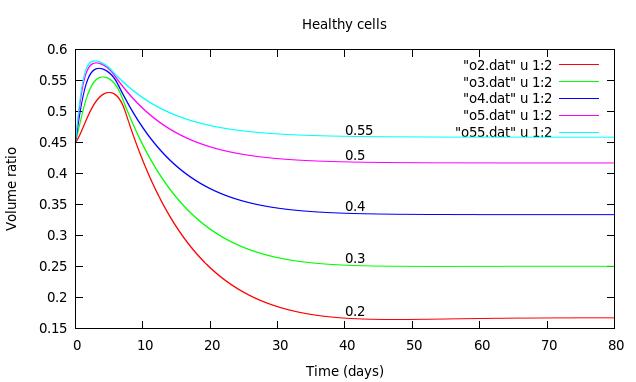} 
  \end{center}
  \caption{Growth of healthy cells varing the oxygen concentration.}
\end{figure}

\begin{figure}[h]
  \begin{center}
    \includegraphics[width=4.05in]{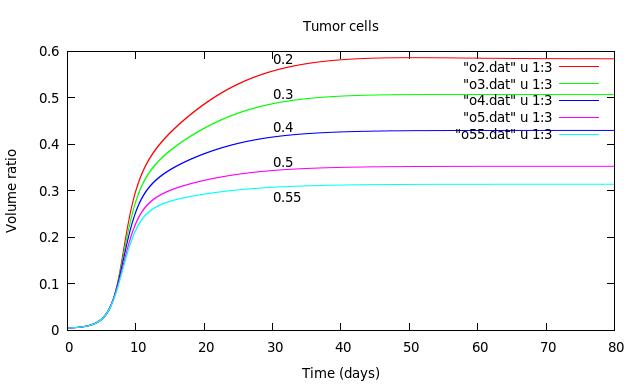} 
  \end{center}
  \caption{Growth of tumor cells varing the oxygen concentration.}
\end{figure}

{\newpage}The parameters used in the simulation remain constant. But if we
consider the parameters of the source for oxygen and glucose (which are
associated with pH levels), these parameters may vary depending on the diet in
each person.\\
For this situation we use uniform random numbers for these parameters, only in the
interval [0, 0.5].

\section{Conclusions}

From the numerical results that we get can say the following: the model
presented in this paper is congruent with Gompertz's model for tumor cells, in
which it is observed a stage of uncontrolled growth and then they stabilize at
a value, also can be seen in graphics that they are susceptible to the
presence of nutrients, these are two cases, if the absorption rate is low the
occupied volume increases and if absorption rate is high the occupied volume
is less. With respect to the variation in the oxygen concentration, can be
seen two situations, if the local concentration of oxygen is low, the volume
in stabilizing the tumor cells is high and when the local concentration of
oxygen is high the volume of tumor cells is lower compared to the previous
situation, this suggests that can intervene in tumor growth increasing the
local concentration of oxygen in the system.\\

\section*{Acknowledgments}
We are grateful for the facilities provided by the Laboratorio Nacional de Superc\'omputo (LNS) del Sureste de M\'exico to obtain these results.

\bibliographystyle{unsrt}
\bibliography{Cancer}

\end{document}